\documentclass[aps,prl,twocolumn,superscriptaddress,showpacs,showkeys]{revtex4}
\begin{document}

\ \\[-4.cm]

\begin{flushright}
NIKHEF 04-024
\end{flushright}

\vspace{-1.5cm}

\title{Nonlinear Bogolyubov-Valatin transformations and quaternions}

\author{J.-W.\ van Holten}\email{v.holten@nikhef.nl}
 \affiliation{NIKHEF, P.O.\ Box 41882, 1009 DB Amsterdam,
The Netherlands}
 \affiliation{Vrije Universiteit Amsterdam,
Faculty of Exact Sciences, Division of Physics and Astronomy,
Department of Theoretical Physics,
De Boelelaan 1081, 1081 HV Amsterdam, The Netherlands}
\author{K.\ Scharnhorst}\email{scharnh@nat.vu.nl}
 \affiliation{Vrije Universiteit Amsterdam,
Faculty of Exact Sciences, Division of Physics and Astronomy,
Department of Theoretical Physics,
De Boelelaan 1081, 1081 HV Amsterdam, The Netherlands}


\begin{abstract}
In introducing second quantization for fermions, Jordan and Wigner (1927/1928)
observed that the algebra of a single pair of fermion creation and
annihilation operators in quantum mechanics is 
closely related to the algebra of quaternions 
{\bf H}. For the first time, here we exploit this fact to study
nonlinear Bogolyubov-Valatin transformations (canonical transformations
for fermions) for a single fermionic mode. By means of these transformations,
a class of fermionic Hamiltonians in an external field is related to
the standard Fermi oscillator.
\end{abstract}



\maketitle

Unitary transformations play a prominent role in quantum mechanics.
Like  canonical transformations in classical mechanics, unitary
transformations of quantum dynamical degrees of freedom often
simplify the dynamical equations, or allow to introduce sensible
approximation schemes. Such methods have wide-ranging applications,
from the study of simple systems to many-body problems in
solid-state or nuclear physics and quantum chemistry, up to
the infinite-dimensional systems of quantum field theory
\cite{blai1,wagn1,ring1,bish1}.
Linear (unitary) canonical transformations 
(i.e., transformations preserving 
the canonical anticommutation relations (CAR))
for fermions have been introduced by Bogolyubov and Valatin 
(for two fermionic modes) 
in connection with the study of the 
mechanism of superconductivity 
\cite{bogo1,pine1,bogo2,vala1,bogo9}.
These (linear) Bogolyubov-Valatin 
transformations have been extended, initially by Bogolyubov and 
his collaborators \cite{bogo3,bogo5}, \cite{bogo6a},
Appendix II, p.\ 123 (\cite{bogo6b}, p.\ 116, \cite{bogo6c}, p.\ 679), 
to involve $n$
fermionic modes [so-called generalized linear Bogolyubov-Valatin
transformations, see, e.g., \cite{bogo11}, Part III, p.\ 247 
(\cite{bogo11b}, p.\ 341)]. Such linear canonical transformations
are important from a physical as well as from a mathematical
point of view. Mathematically, they allow to relate quite
arbitrary Hamiltonians quadratic in the fermion creation and
annihilation operators to collections of Fermi oscillators whose
mathematics is very well understood. From a physical point of view,
canonical transformations implement the concept of quasiparticles
in terms of which the physical processes taking place can
be described and understood in an effective and transparent manner.
To apply the powerful tool of canonical transformations
to the physically interesting class of non-quadratic Hamiltonians, 
however, requires to go beyond linear Bogolyubov-Valatin transformations.
Certain aspects of nonlinear 
Bogolyubov-Valatin transformations 
have received some attention over time
\cite{haar1,kuze1,fuku2,fuku3,fuku4,colp1,fuku1,nish1,nish2,zasl1,fuku5,
gunn1,suzu2,ostl1,holt1,nish3,abe1,katr1,caba1,ilie1}
(We disregard here
work done within the framework of the coupled-cluster method (CCM) 
\cite{bish1} which is nonunitary.).
However, a systematic analytic study of 
general (nonlinear) Bogolyubov-Valatin transformations 
has not been undertaken so far. In the present paper,
as a first step towards this goal we are going to investigate
the prototypical case of a single fermionic mode.

Let us consider a pair of fermion creation and annihilation 
operators $\hat{a}^+$, $\hat{a}$. Here, we regard the 
creation operator $\hat{a}^+$ as the hermitian conjugate of the 
annihilation operator $\hat{a}$: $\hat{a}^+ = \hat{a}^\dagger$ 
(we will use the latter notation throughout). They obey 
the CAR
\begin{eqnarray}
\label{p1a}
\{\hat{a}^\dagger,\hat{a}\}&=&\hat{a}^\dagger\ \hat{a}\ +\
\hat{a}\ \hat{a}^\dagger\ =\ 1, \\
\label{p1b}
\left(\hat{a}^\dagger\right)^2&=&\hat{a}^2\ =\ 0.
\end{eqnarray}
It is now instructive to consider the following 
pair of anti-hermitian operators.
\begin{eqnarray}
\label{p2a}
\hat{a}^{[1]}&=&- \hat{a}^{[1]\dagger}\ =
i\left(\hat{a}+\hat{a}^\dagger\right)\\
\label{p2b}
\hat{a}^{[2]}&=&- \hat{a}^{ [2]\dagger}\ =
\hat{a}-\hat{a}^\dagger
\end{eqnarray}
These two operators obey the equation ($p,q = 1,2$)
\begin{eqnarray}
\label{p3}
\{\hat{a}^{[p]},\hat{a}^{[q]}\}&=&-2\delta_{pq}\ .
\end{eqnarray}
Consequently, they generate the (real) Clifford algebra $C(0,2)$ which
is isomorphic to the algebra of quaternions $\bf H$ (cf., e.g.,
\cite{port2}, Chap.\ 15, p.\ 123, \cite{loun}, Chap.\ 16, p.\ 205).
We can define the three quaternionic units $\bf i$,
$\bf j$, $\bf k$ by the equations
\begin{eqnarray}
\label{p4a}
{\bf i}&=&\hat{a}^{[1]} \ =\ i\left(\hat{a}+\hat{a}^\dagger\right)\ ,\\
\label{p4b}
{\bf j}&=&\hat{a}^{[2]}\ = \ \hat{a}-\hat{a}^\dagger\ ,\\
\label{p4c}
{\bf k}&=&\hat{a}^{[3]}\ =\ \hat{a}^{[1]}\ \hat{a}^{[2]}\ =\
i\left(\hat{a}^\dagger \hat{a} - \hat{a} \hat{a}^\dagger\right)\ .
\end{eqnarray}
Quite generally, these definitions entail that any pair of
fermionic creation and
annihilation operators $\hat{a}^\dagger$, $\hat{a}$ induces
a (bi-)quaternionic structure into any consideration and model they
are a part of. And in turn, any quaternionic structure can be
interpreted in terms of fermionic creation and
annihilation operators. The link between the algebra of quaternions
$\bf H$ and fermion creation and annihilation operators has been
mentioned for the first time by Jordan and Wigner in introducing
second quantization for fermions \cite{jord2}, p.\ 474, \cite{jord1},
p.\ 635 (\cite{schw1}, p.\ 45, \cite{jord3}, p.\ 113).
However, it seems to not have found 
its way into the work of later authors
(The only further mention of this fact in the literature we have been
able to find is in ref.\ \cite{kenn1}.).

Let us now start by writing down an Ansatz for the most general  
Bogolyubov-Valatin transformation for a single fermionic mode.
In view of the eqs.\ (\ref{p1a}), (\ref{p1b}) the new pair of
fermion annihilation and creation operators $\hat{b}$, $\hat{b}^\dagger$ reads
(here we assume the coefficients to be complex numbers: 
$\lambda^{(k;l)}\in{\bf C}$, $k,l=0,1$;
$\{\lambda\} = \{\lambda^{(0;0)},\lambda^{(0;1)},
\lambda^{(1;0)},\lambda^{(1;1)}\}$)
\begin{eqnarray}
\label{p5a}
\hat{b}&=&  {\sf B}\left(\{\lambda\};
\hat{a}\right)\nonumber\\
&=&\lambda^{(0;0)}\; +\;
\lambda^{(0;1)}\ \hat{a}\; +\;
\lambda^{(1;0)}\ \hat{a}^\dagger\; +\;
\lambda^{(1;1)}\ \hat{a}^\dagger\hat{a},\ \\[0.3cm]
\label{p5b}
\hat{b}^\dagger&=& {\sf B}\left(\{\lambda\};
\hat{a}\right)^\dagger\nonumber\\
&=&\overline{\lambda^{(0;0)}}\; +\;
\overline{\lambda^{(0;1)}}\ \hat{a}^\dagger\; +\;
\overline{\lambda^{(1;0)}}\ \hat{a}\; +\;
\overline{\lambda^{(1;1)}}\ \hat{a}^\dagger\hat{a}.\ \
\end{eqnarray}
 From eq.\ (\ref{p1a}) applied to $\hat{b}$, $\hat{b}^\dagger$
follows:
\begin{eqnarray}
\label{p6a}
2\vert\lambda^{(0;0)}\vert^2\ +\ \vert\lambda^{(1;0)}\vert^2\ +\
\vert\lambda^{(0;1)}\vert^2&=&1
\end{eqnarray}
and from eq.\ (\ref{p1b}) follows:
\begin{eqnarray}
\label{p6b}
\left(\lambda^{(0;0)}\right)^2\ +\ \lambda^{(1;0)}
\lambda^{(0;1)}&=&0
\end{eqnarray}
while both equations (\ref{p1a}), (\ref{p1b}) also yield
\begin{eqnarray}
\label{p6c}
2 \lambda^{(0;0)}\ +\ \lambda^{(1;1)}&=&0
\end{eqnarray}
(\cite{holt1}, Sect.\ 2.6, p.\ 32, eq.\ (2.91)).
Using eq.\ (\ref{p6b}), eq.\ (\ref{p6a})
can be transformed to read 
\begin{eqnarray}
\label{p6d}
\vert\lambda^{(1;0)}\vert +\ \vert\lambda^{(0;1)}\vert&=&1
\end{eqnarray}
(Take absolute values on both sides of
the modified eq.\ (\ref{p6b}): 
$\left(\lambda^{(0;0)}\right)^2 = - \lambda^{(1;0)} \lambda^{(0;1)}$, 
eliminate $\vert\lambda^{(0;0)}\vert^2$
from eq.\ (\ref{p6a}) and take the square root.).
For comparison, let us have a look at the class of generalized linear 
Bogolyubov-Valatin transformations (for one mode!):
$\lambda^{(0;0)} = \lambda^{(1;1)} = 0$.
Then, eq.\ (\ref{p6b}) requires that 
$\lambda^{(1;0)}\lambda^{(0;1)} = 0$.
This condition allows two solutions:
\begin{eqnarray}
\label{p7a}
\lambda^{(1;0)}&=&0,\ \ \ \vert\lambda^{(0;1)}\vert\ =\ 1,\\
\label{p7b}
\lambda^{(0;1)}&=&0,\ \ \ \vert\lambda^{(1;0)}\vert\ =\ 1.
\end{eqnarray}
It has been found that generalized
linear Bogolyubov-Valatin transformations (for $n$ modes)
are equivalent to the group of
$O(2n,{\bf R})$ transformations which is in
accord (for $n=1$)
with the eqs.\ (\ref{p7a}), (\ref{p7b}) (This group 
is reduced to $SO(2n,{\bf R})$
if one only allows transformations
continuously connected
to the identity map -- then in our case only eq.\ (\ref{p7a}) applies;
\cite{bloc1,bose1,ixar1,bali1,navo1,beck1}, \cite{fuku1,broa1}, \cite{ohnu1},
Sect.\ 3.2, p.\ 16, \cite{blai1}, Sect.\
2.2, p.\ 38,
\cite{pere1}, Sect.\ 9.1, p.\ 111,
\cite{pere2}, \S 9.1, p.\ 127; if one does
not assume that $\hat{a}$ and $\hat{a}^+$ hermitian conjugates of each other
the corresponding groups are $O(2n,{\bf C})$ and $SO(2n,{\bf C})$,
respectively \cite{bali1,navo1,navo2,beck2}
\cite{blai1}, Sect.\ 2.1, p.\ 34, \cite{zhan1,fan4}.).

The Bogolyubov-Valatin transformation (\ref{p5a}) can be inverted.
We can write:
\begin{eqnarray}
\label{p8}
\hat{a}&=&  {\sf B}\left(\{\nu\};
\hat{b}\right)\nonumber\\
&=&\nu^{(0;0)}\; +\;
\nu^{(0;1)}\ \hat{b}\; +\;
\nu^{(1;0)}\ \hat{b}^\dagger\; +\;
\nu^{(1;1)}\ \hat{b}^\dagger\hat{b}.\ \ 
\end{eqnarray}
Inserting eq.\ (\ref{p5a}) into eq.\ (\ref{p8}) one
obtains a system of linear equations in $\{\nu\}$ whose
(unique) solution reads:
\begin{eqnarray}
\label{p9a}
\nu^{(0;0)}&=&\overline{\lambda^{(0;0)}}\lambda^{(1;0)}\ -\
\lambda^{(0;0)}\overline{\lambda^{(0;1)}}\\
\label{p9b}
\nu^{(0;1)}&=&\overline{\lambda^{(0;1)}}\\
\label{p9c}
\nu^{(1;0)}&=&\lambda^{(1;0)}\\
\label{p9d}
\nu^{(1;1)}&=&-2 \nu^{(0;0)}
\end{eqnarray}
One can convince oneself by explicit calculation that
the $\{\nu\}$ given by eqs.\ (\ref{p9a})-(\ref{p9d})
obey the analogues of eqs.\ (\ref{p6a}), (\ref{p6b})
if the $\{\lambda\}$ obey the latter equations.
Furthermore,
the nonlinear Bogolyubov-Valatin transformations
(\ref{p5a}) form a group $G_{BV}$.
After the above considerations it remains to check that
${\sf B}\left(\{\nu\};\hat{a}\right) =
{\sf B}\left(\{\mu\};{\sf B}\left(\{\lambda\};\hat{a}\right)\right)$
belongs to $G_{BV}$ if ${\sf B}\left(\{\lambda\};\hat{a}\right)$ and
${\sf B}\left(\{\mu\};\hat{a}\right)$ belong to $G_{BV}$.
One can explicitly check that the $\{\nu\}$ obey the
analogues of eqs.\ (\ref{p6a}), (\ref{p6b}) if the $\{\lambda\}$,
$\{\mu\}$ obey the eqs.\ (\ref{p6a}), (\ref{p6b}), or their
analogues, respectively.

To further study the Bogolyubov-Valatin group $G_{BV}$ 
it turns now out to be useful to consider the linear vector 
space $V$ generated by the operators $\hat{a}$, $\hat{a}^\dagger$
($V$ is the space of linear operators in Fock space). It is
four-dimensional and is spanned by the operator basis
$a^T = (1,\hat{a},\hat{a}^\dagger,\hat{a}^\dagger\hat{a})$.
However, taking into account the connection already discussed 
between the operators $\hat{a}$, $\hat{a}^\dagger$ and quaternions
it turns out to be advantageous
to pursue the consideration of this linear space in terms of the
operator basis (cf.\ eqs.\ (\ref{p4a})-(\ref{p4c}))
$a^T = \left(1,\hat{a}^{[1]},\hat{a}^{[2]},\hat{a}^{[3]}\right)$.
The Bogolyubov-Valatin transformation (\ref{p5a}) can be understood
as a base transformation in the linear space $V$. We can write
($b^T = (1,\hat{b}^{[1]},\hat{b}^{[2]},\hat{b}^{[3]})$)
\begin{eqnarray}
\label{p10}
b&=& {\cal A}\left(\{\lambda\}\right)\ a,
\end{eqnarray}
where the $4\times 4$ matrix ${\cal A}\left(\{\lambda\}\right)$ is a block
diagonal matrix 
${\cal A} = {\rm diag}(1,A)$
and $A=A\left(\{\lambda\}\right)$ is the real $3\times 3$ matrix
\begin{eqnarray}
\label{p11}
A\left(\{\lambda\}\right)&=&
%
\left(
\begin{array}{*{3}{c}}
{\rm Re}\;\kappa^{(0;1)}&
{\rm Re}\;\kappa^{(1;0)}&
{\rm Re}\;\kappa^{(1;1)}\\
{\rm Im}\;\kappa^{(0;1)}&
{\rm Im}\;\kappa^{(1;0)}&
{\rm Im}\;\kappa^{(1;1)}\\
{\rm Im}\left(\overline{\kappa^{(1;0)}}\right.&
{\rm Im}\left(\overline{\kappa^{(1;1)}}\right.&
{\rm Im}\left(\overline{\kappa^{(0;1)}}\right.\\
\times\ \kappa^{(1;1)}\Big)&
\times\ \kappa^{(0;1)}\Big)&
\times\ \kappa^{(1;0)}\Big)\end{array}
\right)\ \ \ \ \ 
\end{eqnarray}
with (taking into account eqs.\ (\ref{p6a})-(\ref{p6c}))
unit determinant ($\det A = 1$)
and inverse $A\left(\{\lambda\}\right)^{-1}$ $ = 
A\left(\{\lambda\}\right)^{T}$. Here, we have applied the notation:
\begin{eqnarray}
\label{p12b}
\kappa^{(0;1)}&=&\lambda^{(0;1)}\ +\ \lambda^{(1;0)},\\
\label{p126c}
\kappa^{(1;0)}&=&i\left(\lambda^{(0;1)}\ -\ \lambda^{(1;0)}\right),\\
\label{p12d}
\kappa^{(1;1)}&=&\lambda^{(1;1)}
\ =\ - 2 \lambda^{(0;0)}\ =\ - 2 \kappa^{(0;0)}.
\end{eqnarray}
In view of the above considerations
the Bogolyubov-Valatin group $G_{BV}$
is equivalent to the group $SO(3)$. Given the link 
between creation and annihilation operators and the algebra of
quaternions ${\bf H}$ discussed further above this does not come
as a big surprise. In accordance with eqs.\ (\ref{p4a})-(\ref{p4c}),
the new pair of operators $\hat{b}$, $\hat{b}^\dagger$
defines a transformed system of
quaternionic units  ${\bf i}^\prime$,
${\bf j}^\prime$, ${\bf k}^\prime$ by writing
\begin{eqnarray}
\label{p13a}
{\bf i}^\prime&=&\hat{b}^{[1]}\ =\
i\left(\hat{b}+\hat{b}^\dagger\right)\nonumber\\
&=&{\rm Re}\;\kappa^{(0;1)}\ {\bf i}\; +\;
{\rm Re}\;\kappa^{(1;0)}\ {\bf j}\;
+\; {\rm Re}\;\kappa^{(1;1)}\ {\bf k},\\
\label{p13b}
{\bf j}^\prime&=&\hat{b}^{[2]}\ =\
\hat{b}-\hat{b}^\dagger\nonumber\\
&=& {\rm Im}\;\kappa^{(0;1)}\ {\bf i}\; +\;
{\rm Im}\;\kappa^{(1;0)}\ {\bf j}\;
+\; {\rm Im}\;\kappa^{(1;1)}\ {\bf k},\\
\label{p13c}
{\bf k}^\prime&=&\hat{b}^{[3]}\ =\ \hat{b}^{[1]}\ \hat{b}^{[2]}.
\end{eqnarray}

In terms of the new parameters $\{\kappa\}$ (eqs.\ (\ref{p12b})-(\ref{p12d}))
the equations
(\ref{p6a}), (\ref{p6b}) read
\begin{eqnarray}
\label{p14a}
\vert\kappa^{(0;1)}\vert^2\; +\;
\vert\kappa^{(1;0)}\vert^2\; +\;
\vert\kappa^{(1;1)}\vert^2 &=&2,\\
\label{p14b}
\left(\kappa^{(0;1)}\right)^2\; +\;
\left(\kappa^{(1;0)}\right)^2\; +\;
\left(\kappa^{(1;1)}\right)^2 &=&0.
\end{eqnarray}
Let us now further analyze these equations.
Separating them into real and imaginary parts 
and introducing the three-dimensional (complex) vector
$\left({\bf e}^\prime\right)^T = \left(\kappa^{(0;1)},\kappa^{(1;0)},
\kappa^{(1;1)}\right)$, these (three real)
equations can compactly be written as
\begin{eqnarray}
\label{p15}
\left({\bf e}^\prime\right)^T {\bf e}^\prime&=&0,\ \ 
\vert {\bf e}^\prime\vert^2\ =\ 2.
\end{eqnarray}
${\bf e}^\prime$ is an {\it isotropic vector} (cf., e.g., \cite{altm},
Sect.\ 6.3, p.\ 113, and \cite{codd1} for some more detailed
and pedagogical exposition). In a way, it appears to be an
interesting feature that within the framework of general (nonlinear)
Bogolyubov-Valatin transformations for a single pair of fermion
creation and annihilation operators spinors make their appearance
(via isotropic vectors, cf., e.g., \cite{codd1}). The
properties of these spinors are related  to canonical (Bogolyubov-Valatin)
transformations. Introducing two three-dimensional (real)
vectors 
${\bf e}_1^\prime = {\rm Re}\left({\bf e}^\prime\right) $, 
${\bf e}_2^\prime = {\rm Im}\left({\bf e}^\prime\right)$
(transposed, they agree with the first two rows of the matrix (\ref{p11}))
one can write the eqs.\ (\ref{p15}) as
\begin{eqnarray}
\label{p16}
\vert{\bf e}_1^\prime\vert^2\ =\
\vert{\bf e}_2^\prime\vert^2&=&1,\ \
\left({\bf e}_1^\prime\right)^T {\bf e}_2^\prime\ =\ 0.
\end{eqnarray}
These equations define the vectors
${\bf e}_1^\prime$, ${\bf e}_2^\prime$ as a pair of
orthonormal vectors which can be supplemented by the vector
${\bf e}_3^\prime = {\bf e}_1^\prime\times {\bf e}_2^\prime$
to form an orthonormal vector triple in ${\bf R}_3$. It is worth
mentioning here that the vector $\left({\bf e}_3^\prime\right)^T$
coincides with the third row of the matrix (\ref{p11}).
Consequently, the orthogonality condition(s) for the matrix
(\ref{p11}) are equivalent to the conditions for the
Bogolyubov-Valatin transformation to be canonical (eqs.\
(\ref{p14a}), (\ref{p14b}) or (\ref{p6a}), (\ref{p6b})). 
This is a generalization
of an insight obtained for linear Bogolyubov-Valatin transformation
(see \cite{bloc1,bali1,beck1}) to the general (nonlinear) case.

The canonical (Bogolyubov-Valatin) transformation (\ref{p5a}) 
can be implemented by means of an unitary transformation
${\sf U}\left(\{\lambda\};\hat{a}\right)$:
\begin{eqnarray}
\label{p17}
\hat{b}&=& {\sf B}\left(\{\lambda\};\hat{a}\right)
\ =\ {\sf U}\left(\{\lambda\};\hat{a}\right)\ \hat{a}\
{\sf U}\left(\{\lambda\};\hat{a}\right)^\dagger.
\end{eqnarray}
The analogue of eq.\ (\ref{p17})
\begin{eqnarray}
\label{p18}
\hat{b}^{[1]}&=&
\ =\ {\sf U}\left(\{\lambda\};\hat{a}\right)\ \hat{a}^{[1]}\
{\sf U}\left(\{\lambda\};\hat{a}\right)^\dagger
\end{eqnarray}
has a remarkable interpretation in terms of
quaternions discussed further above (an analogous comment
applies to $\hat{a}^{[2]}$ and $\hat{a}^{[3]}$). Eqs.\ (\ref{p13a}) and
(\ref{p18}) are just concrete realizations of the theory of rotations
in the language of quaternions first elaborated by Cayley and
Hamilton
(cf., e.g., \cite{altm}, Sect.\ 12.8, p.\ 215, eq.\ (9),
\cite{bied}, Sect.\ 4.5, p.\ 201). Eq.\ (\ref{p13a})
represents a ($SO(3)$) rotation of the vector $(1,0,0)$
in the three-dimensional space
spanned by the quaternionic units $\bf i$, $\bf j$, $\bf k$ while eq.\
(\ref{p18}) stands for the corresponding ($SU(2)$) transformation
of the quaternion ${\bf i}$ ($= \hat{a}^{[1]}$) by
quaternionic multiplication. The unitary operator
${\sf U}\left(\{\lambda\};\hat{a}\right)$ can be understood as
a unit quaternion given by ($-\pi < \phi\le\pi$, $n_1, n_2, n_3 \in {\bf R}$,
$n^2 = 1$)
\begin{eqnarray}
\label{p19a}
{\sf U}\left(\{\lambda\};\hat{a}\right)&=&
\cos\frac{\phi}{2}\; +\;
\sin\frac{\phi}{2}\ \left(n_1\ {\bf i} + n_2\ {\bf j} + n_3\ {\bf k}\right)
\ \ \ \\
\label{p19b}
&=&{\rm e}^{\displaystyle\ \phi
\left(n_1\ {\bf i} + n_2\ {\bf j} + n_3\ {\bf k}\right)/2}\ .
\end{eqnarray}
The coefficients $\{\lambda\}$ are given in terms of the parameters
$\phi$, $n_1$, $n_2$, $n_3$ by the equations
\begin{eqnarray}
\label{p19c}
\lambda^{(0;1)}&=&
\left(\cos\frac{\phi}{2} - i n_3\; \sin\frac{\phi}{2}\right)^2,\\
\label{p19d}
\lambda^{(1;0)}&=&\left(n_1 + i n_2\right)^2\ \sin^2\frac{\phi}{2},\\
\label{p19e}
\lambda^{(1;1)}&=&- 2 \lambda^{(0;0)}\nonumber\\
&=& 2 i \left(\cos\frac{\phi}{2} - i n_3\; \sin\frac{\phi}{2}\right)
\left(n_1 + i n_2\right)\; \sin\frac{\phi}{2} .\ \ \ \
\end{eqnarray}
To obtain these relations insert eq.\ (\ref{p19a})
into eq.\ (\ref{p17}) and compare the r.h.s.\ with eq. (\ref{p5a}).
From the representation (\ref{p19b}) one sees immediately that the operators
${\bf i} = \hat{a}^{[1]}$, ${\bf j} = \hat{a}^{[2]}$,
${\bf k} = \hat{a}^{[3]}$ (cf.\ eqs.\ (\ref{p4a})-(\ref{p4c}))
are generators of the group $SU(2)$ and they obey the Lie algebra
of  $SO(3)$, $SU(2)$. This has been observed earlier (in a more general
context) in \cite{wybo1} (also see \cite{fuku2}).
Related observations can be found in \cite{bosesk1}, Appendix A.1, 
p.\ 919, \cite{buza1} and \cite{zhan}, p.\ 907, eq.\ (6.2).
One can convince oneself that for linear Bogolyubov-Valatin transformations
of type (\ref{p7a}) eq.\ (\ref{p19b}) agrees (sometimes up
to some elementary complex
phase factor) with eq.\ (7) in \cite{ziet1}, with
eq.\ (5.1) in \cite{beck1}, with eq.\ (3.6) in \cite{beck2},
with eq.\ (2.32a), p.\ 40, Sect.\ 2.2 in \cite{blai1},
with eq.\ (3.10) in \cite{fan3} (reduced to the one-mode case;
incidentally, there is disagreement with \cite{zhan2}, p.\ 205,
below of eq.\ (11)).

The vacuum state $\vert 0\rangle$ defined by
$\hat{a}\vert 0\rangle = 0$
transforms under (general, i.e., $SO(3)$) Bogolyubov-Valatin transformations
according to the law
\begin{eqnarray}
\label{p20}
\vert 0\rangle_{\{\lambda\}}&=&
{\sf U}\left(\{\lambda\};\hat{a}\right)\vert 0\rangle\ ,\ \
\hat{b}\vert 0\rangle_{\{\lambda\}}\ =\ 0.
\end{eqnarray}
Associating $\vert 0\rangle$ with a vector in a two-dimensional (complex)
Hilbert space and ${\sf U}\left(\{\lambda\};\hat{a}\right)$ with a
$2\times 2$ matrix operating in it
[cf.\ \cite{jord2}, p.\ 474/475,
\cite{jord1}, p.\ 634 (\cite{schw1}, p.\ 44, \cite{jord3}, p.\ 112)] 
one sees that this
vector transforms as a spinor (with a corresponding element of $SU(2)$)
under Bogolyubov-Valatin transformations.
The state $\vert 0\rangle_{\{\lambda\}}$
is a spin ($SU(2)$) coherent state \cite{fuku1}
(with respect to the $\hat{a}$, $\hat{a}^\dagger$ operators,
cf., e.g., \cite{klau2}, Sect.\ I.4, p.\ 25, \cite{pere1}, Sect.\ 4.3,
p.\ 59, \cite{pere2}, \S 4.3, p.\ 72, \cite{zhan}, Sect.\
III.D.1, p.\ 884, and Sect.\ VI.A.1, p.\ 907). However, these fermion coherent
states are different (cf.\ the comments in \cite{klau2}, Sect.\ I.5, p.\ 55
and in \cite{zhan}, Sect.\ VI.D,
p.\ 919) from the Grassmann (fermion)
coherent states (see, e.g., \cite{klau2}, Sect.\ I.5, p.\ 48).

Finally, let us have a look at the standard Fermi oscillator given
by the Hamiltonian $H = \hat{a}^\dagger\ \hat{a} - \frac{1}{2}$.
Applying the Bogolyubov-Valatin transformation (\ref{p5a}) one can
see that it is unitarily equivalent to the following
class of fermionic oscillators in an external field 
\cite{haar1,kuze1,fuku2,colp1,zasl1}, \cite{holt1}, Sect.\ 2.6, p.\ 29
(below, we have taken into account the eqs.\ 
(\ref{p6a})-(\ref{p6d})):
\begin{eqnarray}
\label{p21}
H^\prime &=&\hat{b}^\dagger\ \hat{b}\; -\; \frac{1}{2}\ =\ 
{\sf U}\left(\{\lambda\};\hat{a}\right)\ H\
{\sf U}\left(\{\lambda\};\hat{a}\right)^\dagger\nonumber\\
&=&\left(\vert\lambda^{(0;1)}\vert - \vert\lambda^{(1;0)}\vert\right)
\left(\hat{a}^\dagger\ \hat{a}- \frac{1}{2}\right)\nonumber\\
&&+\;\left( \overline{\lambda^{(0;0)}} \lambda^{(0;1)} 
- \lambda^{(0;0)} \overline{\lambda^{(1;0)}}\right) \hat{a}\nonumber\\
&&+\;\left(\lambda^{(0;0)}\overline{\lambda^{(0;1)}} 
- \overline{\lambda^{(0;0)}} \lambda^{(1;0)}\right) \hat{a}^\dagger.
\end{eqnarray}
As a special case, eq.\ (\ref{p21}) contains for 
$\vert\lambda^{(0;1)}\vert = \vert\lambda^{(1;0)}\vert$
also Hamiltonians that are linear in the creation and annihilation
operators (such Hamiltonians have been studied in 
\cite{colp1}, Sect.\ 4, p.\ 477). Eq.\  (\ref{p21}) demonstrates
that any (Hermitian) Hamiltonian $H_0$ ($0\le\alpha\in {\bf R}$, 
$\beta\in {\bf C}$)
\begin{eqnarray}
\label{p22}
H_0 &=&\alpha \left(\hat{a}^\dagger\ \hat{a} - \frac{1}{2}\right)
\ +\ \beta\; \hat{a}\ +\ \overline{\beta}\; \hat{a}^\dagger
\end{eqnarray}
can be written and understood in terms of transformed creation and annihilation
operators $\hat{b}^\dagger$, $\hat{b}$ as
\begin{eqnarray}
\label{p23}
H_0 &=&\sqrt{\alpha^2 + 4\vert\beta\vert^2}\ 
\left(\hat{b}^\dagger\ \hat{b} - \frac{1}{2}\right).
\end{eqnarray}
In view of the above considerations, its dynamical (spectrum
generating) algebra is ${\rm so(3)}\sim {\rm su(2)}$.

The present paper paves the way for the study of (nonlinear) 
Bogolyubov-Valatin transformations in full generality 
for any finite number of fermionic modes. This is done
by introducing a methodological framework which can be
generalized (stepwise) to more than just one mode. 
For several - say $n$ - fermionic modes, the formalism can,
for example, be expected to allow equivalences of wide classes of 
non-quadratic fermionic Hamiltonians to collections of 
$n$ Fermi oscillators to be derived. This will be of considerable
interest for a wide range of physically relevant models. 
However, beyond its plain methodological 
value the present study of nonlinear Bogolyubov-Valatin 
transformations for just one fermionic mode 
provides us even with some surprising insight. Note, that 
the nonlinear Bogolyubov-Valatin transformation (\ref{p5a}) defines
fermionic operators as a sum of fermi-even and fermi-odd terms.
This is reminiscent of a supersymmetric transformation (cf.\ in this
respect \cite{ilie1}). In this context, remember 
that linear Bogolyubov-Valatin 
transformations - in contrast to (bosonic) linear Bogolyubov 
transformations - do not allow any linear shifts by complex numbers
to be performed [see, e.g., \cite{bogo8a}, 
Appendix IV, Sect.\ 1(a), 1.\ ed.: p.\ 280,
2.\ ed.: p.\ 292 (\cite{bogo8b}, p.\ 328), 
also see \cite{blai1}, Sect.\ 2.4, p.\ 40, eq.\ (2.33b)].

The future generalization of the present work to several fermionic
modes can alternatively also be understood as generalizing the 
complex coefficients $\{\lambda\}$ in eq.\ (\ref{p5a}) to 
operator valued functions for which the present analysis has to 
be repeated in an appropriately modified manner. 
Closely related to this direction of future research is
to consider the coefficients $\{\lambda\}$ as elements of 
an appropriately chosen Grassmann algebra.
However, the coefficients $\{\lambda\}$ can not
only be imagined to be functions of fermionic operators but also
to be functions of bosonic creation and annihilation operators. Such
constructions are met, for example, in the study of so-called
quantized Bogolyubov-Valatin transformations (introduced in \cite{suzu1})
and, more generally, in the study of boson-fermion interactions
(see, e.g., \cite{wagn1}, Chap.\ 5, p.\ 108).

\end{document}